\documentclass[intlimits,twoside,a4paper]{article}

\usepackage{amsmath,amssymb}
\usepackage{graphicx}
\usepackage{wrapfig}

\usepackage[T2A]{fontenc}
\usepackage[cp1251]{inputenc}
\usepackage{cmpj2}

\issue{2012}{15}{1}{13601}

\doinumber{10.5488/CMP.15.13601}

\newcommand{\eps}{\varepsilon}

\title[Rochelle salt: the role of
thermal strains]%
{Major physical characteristics of Rochelle salt: \\the role of
thermal strains}
\author[A.P. Moina]{A.P. Moina}

\authorcopyright{A.P. Moina, 2012}

\address{Institute for Condensed Matter Physics
of the National Academy of Sciences of Ukraine,\\
1 Svientsitskii Str., 79011 Lviv, Ukraine}

\date{Received August 29, 2011, in final form November 22, 2011}

\begin{document}

\maketitle

\begin{abstract}
We compare the results for the related to the shear strain
$\eps_4$ physical characteristics of Rochelle salt obtained within
the recently developed modified two-sublattice Mitsui model that
takes into account the strain $\eps_4$ and the diagonal components
of the strain tensor $\eps_1$, $\eps_2$, $\eps_3$ with the results
of the previous modification of the Mitsui model with the strain
$\eps_4$ only. Within the framework of the model with the diagonal
(thermal expansion) strains, we also reexamine the effects of the
longitudinal electric field $E_1$ on the dielectric properties of
Rochelle salt.

\keywords Rochelle salt, thermal expansion, Mitsui model, strains,
 bias field
\pacs 65.40.De, 77.65.Bn, 77.22.Gm, 77.22.Ch
\end{abstract}

\section{Introduction}
Rochelle salt is the first known ferroelectric and a curious
material, undergoing two second-order phase transitions at $T_{\rm
C1}=255$~K and $T_{\rm C2}=297$~K, with the intermediate
ferroelectric phase.  Spontaneous polarization $P_1$ is directed
along the $a$ axis and accompanied by spontaneous shear strain
$\varepsilon_4$ in the $bc$ plane.

Microscopic theories of Rochelle salt are usually based on the
two-sublattice Mitsui model~\cite{80}, which considers motion of
certain ordering units in two interpenetrating sublattices of
asymmetric double-well potentials. Deformational effects have to
be included into the Mitsui model to get a proper description of
Rochelle salt behavior. Thus, only with taking into account the
piezoelectric coupling with $\eps_4$~\cite{ourrs} it yields the
qualitatively correct behavior of the relaxation time and dynamic
dielectric permittivity near the Curie temperatures. The same
modification of the Mitsui model also provides a fair description
of the major dielectric, piezoelectric, and elastic
characteristics of Rochelle salt associated with the strain
$\eps_4$~\cite{ourrs}, although some systematic discrepancies
between the theory and experiment do take place.

Recently, a further modification of the Mitsui model has been
proposed~\cite{our-diagonal} in order to consistently describe a
number of effects associated with the diagonal components of the
lattice strain tensor $\eps_i$ ($i=1,2,3$). Along with the shear
strain $\eps_4$ included into the Mitsui model in~\cite{ourrs},
this modification takes into account the diagonal strains in the
manner suggested in~\cite{monoclinic} as well as the host lattice
contribution into the energy of thermal expansion. It has been
shown~\cite{our-diagonal} to be efficient in describing the
effect of hydrostatic, uniaxial, or biaxial pressure applied
along the orthorhombic crystallographic axes of the crystal, its
thermal expansion, as well as related to the diagonal strains
components of piezoelectric (e.g. $d_{1i}$, $i=1,2,3$) and elastic
($c_{i4}$) tensors, appearing in the ferroelectric phase only due
to the lowered crystal symmetry.

In the present paper we shall explore how the inclusion of
thermal strains affect the agreement with experiment for the
physical characteristics of Rochelle salt related to the shear
strain $\eps_4$.

It has been  also shown~\cite{our-field} that to correctly describe
 the dependences of the static dielectric permittivity of
Rochelle salt on the longitudinal electric field $E_1$ at
temperatures near $T_{\rm C2}$ (within the modified Mitsui model
without the diagonal strains~\cite{ourrs}), one has to assume that
the external field is partially screened out due to the
space-charge build-up at blocking electrodes. We shall verify
whether this assumption is necessary for the model with the
diagonal (thermal) strains~\cite{our-diagonal}.

\section{$\eps_4$-strain associated physical characteristics in presence \\of diagonal strains}
The expressions for thermodynamic potential, polarization and
strains, as well as for the observable physical characteristics --
second derivatives of thermodynamic potential, obtained within
the Mitsui model with the diagonal strains, can be found elsewhere
\cite{our-diagonal}. In the present paper we shall be interested
only in the major characteristics, associated with polarization
$P_1$ and shear strain $\eps_4$. Those will be compared with the
results obtained within the Mitsui model without diagonal
strains~\cite{ourrs}.

The characteristics we are interested in are spontaneous
polarization $P_1$ and strain $\eps_4$, dielectric susceptibility
of a clamped crystal $\chi_{11}^{\varepsilon}$, and the elastic
constant at a constant field $c_{44}^E$, the expressions for which
are not modified by the inclusion of diagonal strains, except for
renormalization of the model parameters (see~\cite{ourrs} and
\cite{our-diagonal}).
 On the other hand, the static dielectric susceptibility of a mechanically free
crystal $\chi_{11}^{\sigma}$~\cite{our-diagonal,monoclinic} and
the piezoelectric constant
\begin{equation}
d_{14}= d_{14}^0-
 \frac{s_{44}^{E0}\mu_1' \beta \psi_4}{v} \frac{\varphi_3}
{\varphi_2-\Lambda\varphi_3}+\sum_{j=1}^3s_{j4}^E e_{1j}
\end{equation}
contain new terms (the last sum), describing the contribution of
the diagonal strains. Here  $e_{1i}$ are monoclinic
piezoelectric coefficients different from zero only at non-zero
polarization;  $s_{ij}^E$ is the matrix of elastic compliances,
inverse to the matrix of elastic constants $c_{ij}^E$, the
microscopic expressions for which, as well as the notations
introduced above, have been presented earlier~\cite{our-diagonal}.
In the paraelectric phases the expressions for free
susceptibility and $d_{14}$ coincide with those obtained within
the modified Mitsui model without thermal strains
\cite{ourrs2}.

The dynamics of Rochelle salt in the presence of diagonal strains is explored~\cite{6a} elsewhere. The
dynamic dielectric permittivity of a clamped crystal  relevant to the present consideration is not
explicitly dependent on the strains, apart from renormalization of
the interaction constants. The expression for it is the same as in
the model without thermal strains~\cite{ourrs,ourrs2}
\begin{equation} \label{chi_clamped}
\eps_{11}^\eps(\omega)=\eps_{11}^{\eps0}+\frac{\beta\mu_1^2}{2v\eps_0}F_1(\alpha\omega),\qquad
F_1(\alpha\omega)=\frac{\ri\alpha\omega \lambda_1 +
\varphi_3}{(\ri\alpha\omega)^2 + (\ri\alpha\omega) \varphi_1
+\varphi_2 }\,,\end{equation}
 where
$\alpha$  is the parameter setting the time scale of the dynamic
processes in the pseudospin subsystem within the Glauber approach;
the other notations are given in~\cite{ourrs2}.

\section{Numerical calculations}

\subsection{Model parameters}

The values of the parameters of the model with diagonal strains
were chosen~\cite{our-diagonal} to provide as good as possible fit
of the theory to the experimental data for the following
characteristics of Rochelle salt: the Curie temperatures at
ambient pressure $T_{{\rm C}k}$ ($k=1,2$) and their hydrostatic
and uniaxial pressure slopes, the temperature curves of thermal
expansion strains $\eps_i$, linear thermal expansion coefficients,
monoclinic piezoelectric coefficients, and elastic constants
$c_{ij}$ and $c_{i4}$ ($i,j=1,3$).
 The adopted values of the model
parameters  and the details of the fitting procedure have been
given elsewhere~\cite{our-diagonal}.

Additionally, we need to determine the value of the parameter
$\alpha$ that sets the time scale of the model pseudospin
dynamics. In earlier calculations~\cite{ourrs} it was taken to be
temperature independent $\alpha=1.7\cdot 10^{-13}$~s. However, the
analysis~\cite{35} of the temperature dependences of the
relaxation time $\tau$ and of the low-frequency limit of the
clamped dynamic permittivity (measured at 155 MHz, just above the
resonances, the experimental analog of the static clamped
permittivity $\eps^\eps$) revealed that $\tau$ is proportional to
$\eps^\eps-\eps_{\infty}$ and to $T^N$, with $N=1.25\pm0.25$. In
terms of our model, it means that we should take $\alpha$ to be
temperature dependent. The best results are obtained at $
\alpha=\alpha_0\left({T}/{T_{\rm C2}}\right)^{1.35}, $ with
$\alpha_0=2.15\cdot 10^{-13}$~s.

The adopted set of the model parameters is not unique; there are
many other sets providing the fit to experimental data with the
same error. Therefore, it is not possible to precisely establish the
temperature variation of the interaction constants.
However, the overall tendency is such that an increasing temperature
decreases the asymmetry parameter $\Delta$ and the constants of
interactions between the pseudospins within the same and in
different sublattices $J$ and $K$. In other words, increasing
temperature has the effect opposite to hydrostatic compression~\cite{our-diagonal},
which is understandable.  The corresponding temperature slopes at
the adopted values of the model parameters are  $\partial \ln
\Delta/\partial T=-2\cdot 10^{-2}$ K$^{-1}$, $\partial \ln
J/\partial T=-6\cdot 10^{-4}$ K$^{-1}$, and $\partial \ln
K/\partial T=-2\cdot 10^{-2}$ K$^{-1}$.

The number and (if any) temperature and order of the phase
transitions for the Mitsui model without thermal strains are
usually analyzed in terms of dimensionless variables $\bar a$
and $\bar b$
\begin{equation}
 \bar a = \frac{ K -  J}{K +  J +
\frac{8}{v}\psi_4^2s_{44}^{E0}}\,, \qquad  \label{dimensionless} \bar
b = \frac{8 \Delta}{ K + J + \frac{8}{ v}\psi_4^2s_{44}^{E0}}\,.
 \end{equation}
Here, $\psi_4$ is the parameter describing the coupling between the
pseudospin subsystem and the shear strain $\eps_4$~\cite{ourrs}.
The phase diagram of the conventional (undeformable) Mitsui model
in the $(\bar a, \bar b)$ plane~\cite{vaks,werch,Dublenych} shows
the regions with different numbers and types of phase
transitions; its topology does not change by inclusion of the shear
strain $\eps_4$.

\begin{wrapfigure}{i}{0.55\textwidth}
\centerline{\includegraphics[width=0.45\textwidth]{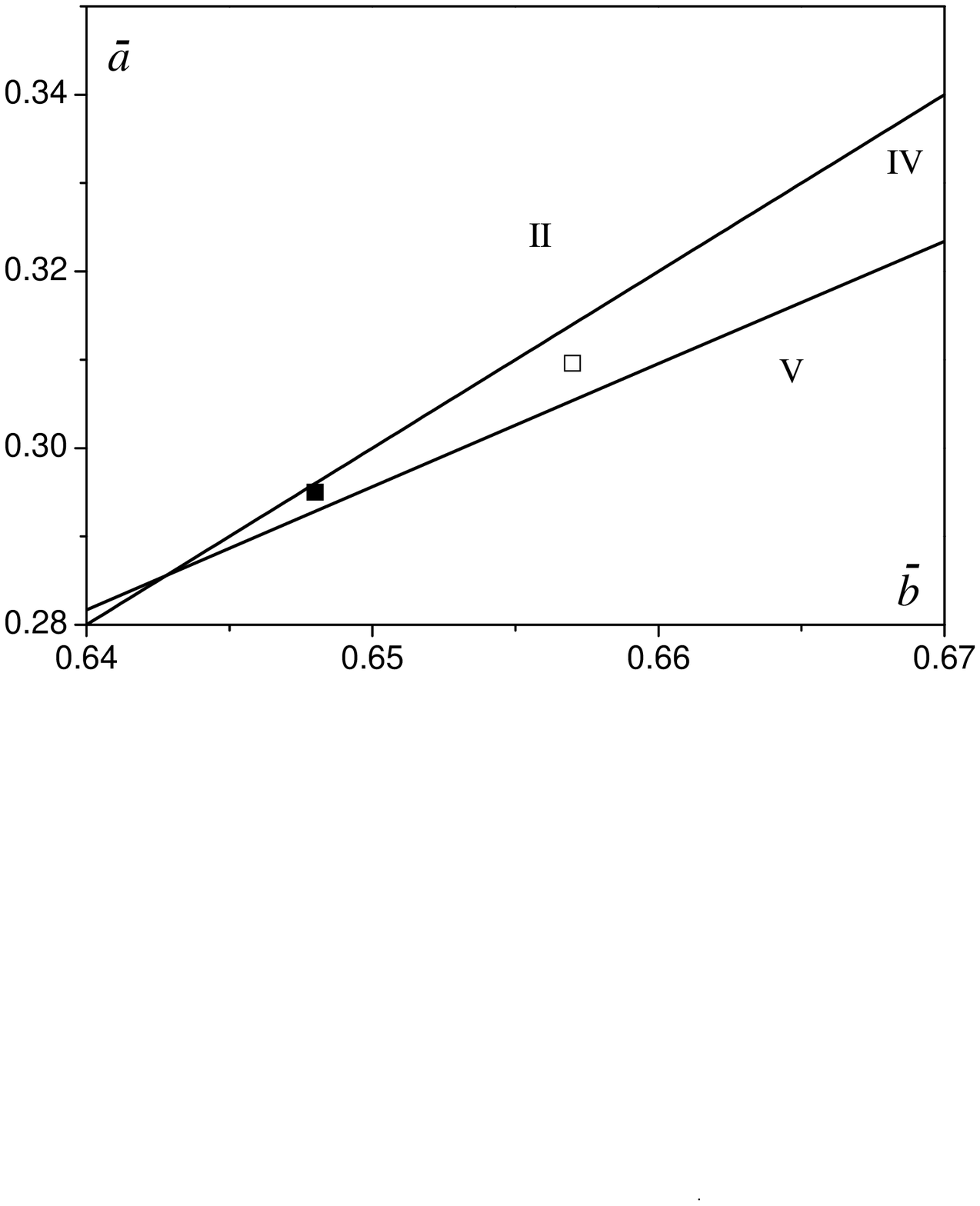}}
\caption{Part of the phase diagram of the Mitsui model. The points
corresponding to the parameters adopted in~\cite{ourrs}
($\blacksquare$) and~\cite{our-diagonal} ($\square$) are shown.}
\label{fig-pd}
\end{wrapfigure}

It has been found that only in a very narrow region of the $(\bar
a, \bar b)$ plane, the system undergoes two second order phase
transitions with the intermediate ferroelectric phase; this region
(IV) is shown in figure~\ref{fig-pd}. In the region V, the system
undergoes no phase transition, whereas in the region II the system
undergoes two second order phase transitions with the
intermediate ferroelectric phase and below them~--- an additional
first order phase transition to another ferroelectric phase, which
persists down to 0~K.

The situation drastically changes in presence of diagonal strains,
when $\bar a$ and $\bar b$ become functions of temperature and
pressure. Here, the number and type(s) of phase transitions are
governed by more than 10 model parameters with arbitrary
(physically reasonable) values. At such circumstances we think it
is impossible to construct a 2D or 3D phase diagram of the model,
which would fully describe all the types of the temperature
behavior and phase transitions in the system.

In the fitting procedure we deal with the values of $\bar a$ and
$\bar b$ at the upper Curie temperature and ambient pressure $\bar
a_0$ and $\bar b_0$. The used values of $\bar a_0$ and $\bar b_0$
are from the region IV of the $(\bar a, \bar b)$ phase diagram of
the undeformable Mitsui model. The points ($\bar a$, $\bar
b)=(0.3162, 0.662)$ corresponding to the parameters of~\cite{ourrs} and ($\bar a_0$, $\bar b_0)=(0.295, 0.648)$ for the
parameters of~\cite{our-diagonal} are shown in
figure~\ref{fig-pd}.

 With decreasing $\bar a_0$, the maximal values
of the order parameter $\xi$ and spontaneous strain $\eps_4$
increase. In the previous model~\cite{ourrs} we had to choose, in
fact, the point almost on the boundary of the region IV, providing
the maximal possible values of $\xi$ and $\eps_4$ in the
middle of the ferroelectric phase. Nevertheless, the calculated
spontaneous polarization and spontaneous strain were still
appreciably smaller than the experimental ones. Interestingly, at
the same values of $(a_0,b_0)$ for the model~\cite{our-diagonal}
and  $(a,b)$ for the model~\cite{ourrs}, the maximal value of the
order parameter obtained within the model with the thermal strains
is much larger. Thus, at the adopted values of the
fitting parameters it is $\xi_{\mathrm{max}}=0.144$, to be compared to
0.128 in~\cite{ourrs}.

\subsection{$\varepsilon_4$-strain-related characteristics: role of diagonal strains}

\begin{figure}[hbt]
\centerline{\includegraphics[width=0.35\textwidth]{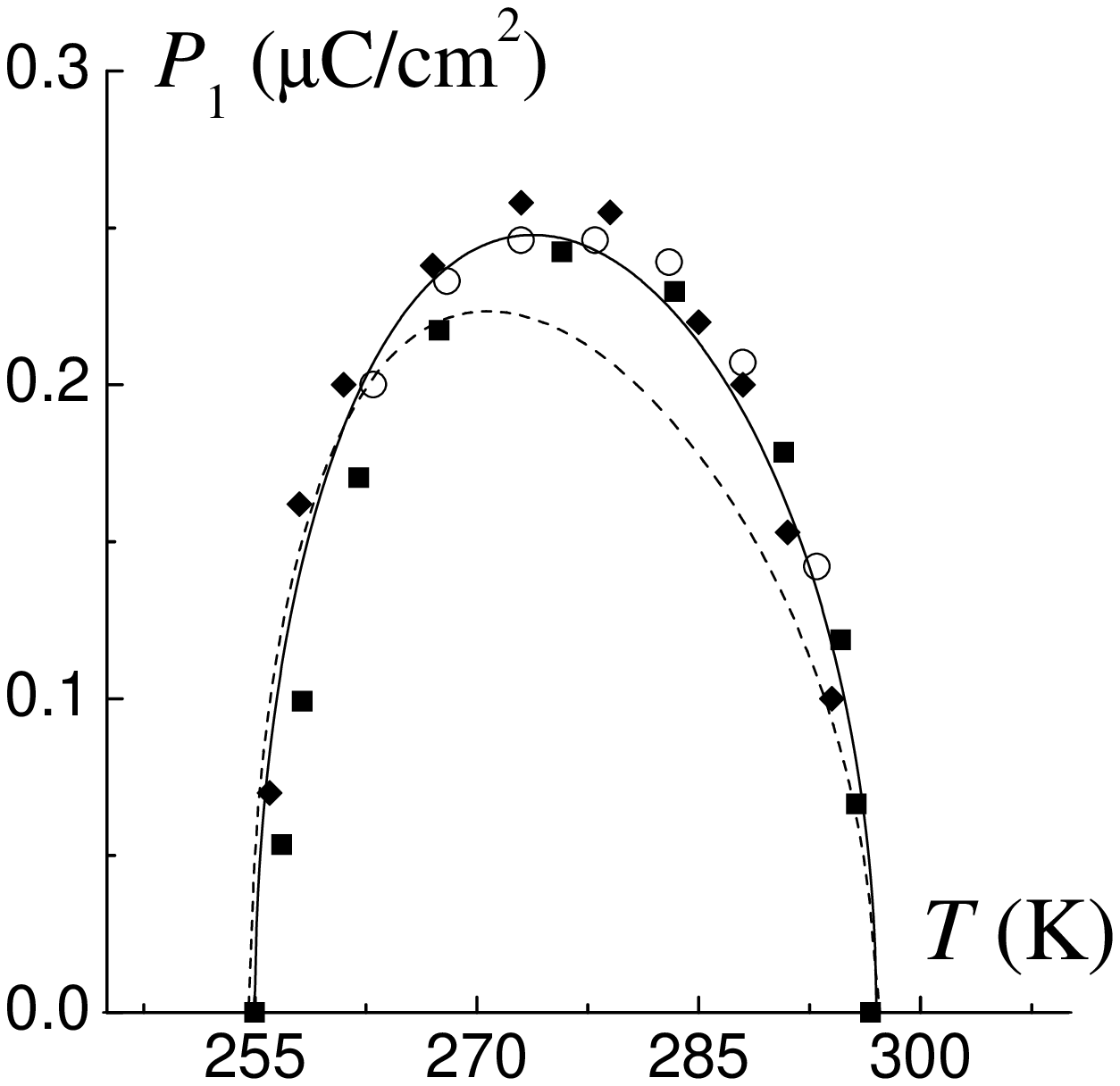}
\includegraphics[width=0.35\textwidth]{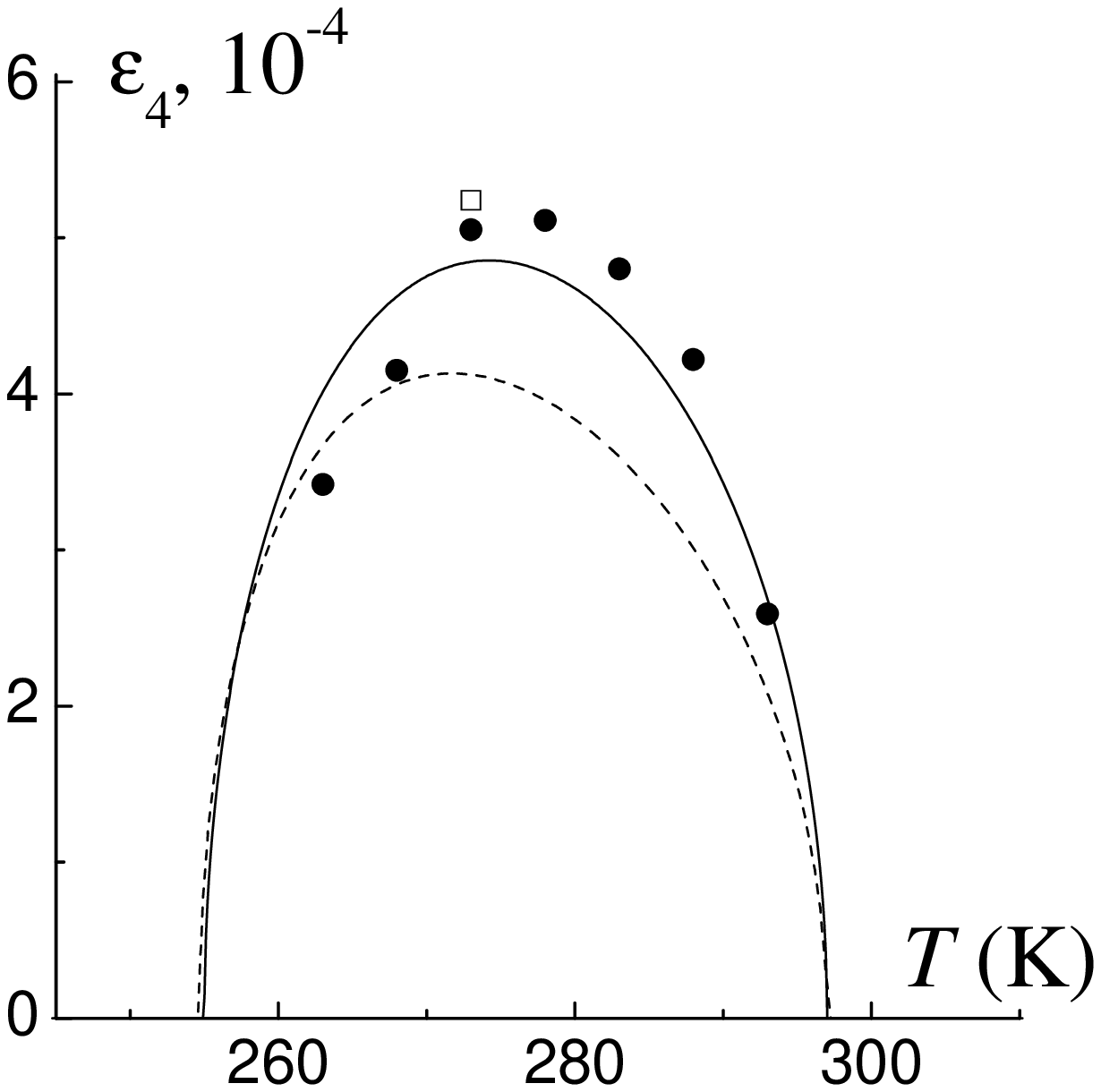}}
\caption{\label{fig-comp-pol} Temperature dependence of
spontaneous polarization (left) and spontaneous strain $\eps_4$
(right) of Rochelle salt: \protect\Large$\circ$\protect\small{}~---~\cite{39};
$\blacksquare$~---~\cite{15}; $\blacklozenge$~---
\cite{24}; $\square$~---~\cite{7}, $\bullet$~---
$\eps_4={P_1d_{14}}/{\chi_{11}^{\sigma}}$~\cite{39}. Lines: a
theory. Solid line: the model of~\cite{our-diagonal}; dashed line:
the model of~\cite{ourrs}. Symbols: experimental points.}
\end{figure}

Figures~\ref{fig-comp-pol} show that the modified Mitsui model
with thermal strains~\cite{our-diagonal} yields a notably larger
values of the spontaneous polarization $P_1$ and spontaneous
strain $\eps_4$ in the middle of the ferroelectric phase and a
better agreement with experimental data. The obtained improvement
is explained by the increased values of the order parameter in the
center of a ferroelectric phase.

The role of diagonal strains is not so apparent for the other physical characteristics of
Rochelle salt
related to the shear strain $\eps_4$. In particular this concerns the static dielectric
permittivities of the free $\eps_{11}^\sigma$ and clamped
$\eps_{11}^\eps$ crystals, as well as the dynamic permittivity.
Here, the differences between the obtained results, with one
exception, are caused by a different route, taken in
\cite{our-diagonal} at selecting the values of the parameters
$\mu_1$ and $\alpha$. The same route could be taken within the
previous modification of the Mitsui model.

\begin{wrapfigure}{i}{0.5\textwidth}
\vspace{-5mm}%
\centerline{\includegraphics[width=0.438\textwidth]{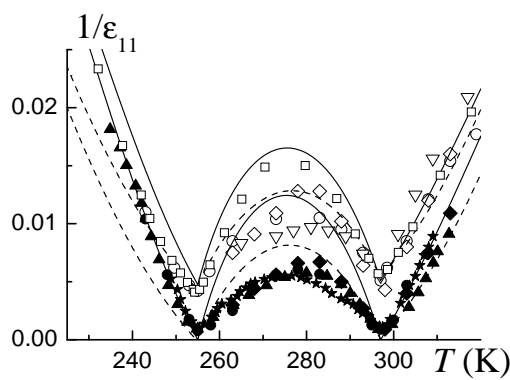}}
\caption{ Temperature dependence of inverse static dielectric
permittivity of a free: $\bigstar$~---~\cite{Lunk};
$\blacktriangle$~---~\cite{29}; $\blacklozenge$~---~\cite{24};
$\bullet$~---~\cite{27} and clamped:~--- $\square$~---~\cite{35};
  \protect\Large$\circ$\protect\small{}~---~\cite{31}; $\Diamond$~---
\cite{24};  $\bigtriangledown$~---~\cite{33} crystals. Lines: the
theory. Solid line: the model of~\cite{our-diagonal}; dashed line:
the model of~\cite{ourrs}. Symbols: experimental points.}
\label{fig-comp-chi}
\vspace{-5mm}%
\end{wrapfigure}

In the previous work~\cite{ourrs} we were not able to simultaneously fit
 the free and clamped dielectric permittivities in
the lower paraelectric phase with the same values of the dipole
moment $\mu_1$ (the dashed lines in figure~\ref{fig-comp-chi}). It
was chosen to provide the best fit for the dynamic dielectric
permittivity~\cite{35} at microwave frequencies and for the
static clamped permittivity at temperatures just below $T_{\rm
C1}$. Thence, the agreement with experiment for the free
permittivity at these temperatures was unsatisfactory. Moreover,
at even lower temperatures (below 240~K) the fit for the dynamic
permittivity was also bad. In the upper paraelectric phase both
free and clamped permittivities were well described.

The problem persists after the inclusion of diagonal strains.
However, now we choose the dipole moment $\mu_1$ to provide the
best fit for the free static permittivity in both paraelectric
phases and for the dynamic dielectric permittivity~\cite{35} below
240~K. As a result, the agreement with experiment for the static
clamped permittivity just below $T_{\rm C1}$ is spoiled. This is
illustrated in figure~\ref{fig-comp-chi}.

In the ferroelectric phase, there is a considerable inconsistency
between the two theories and experiment. This is attributed to the
domain wall contributions into the static free permittivity that
are not taken into account by either modification of the Mitsui
model.

\begin{figure}[hbt]
\centerline{\includegraphics[width=0.49\textwidth]{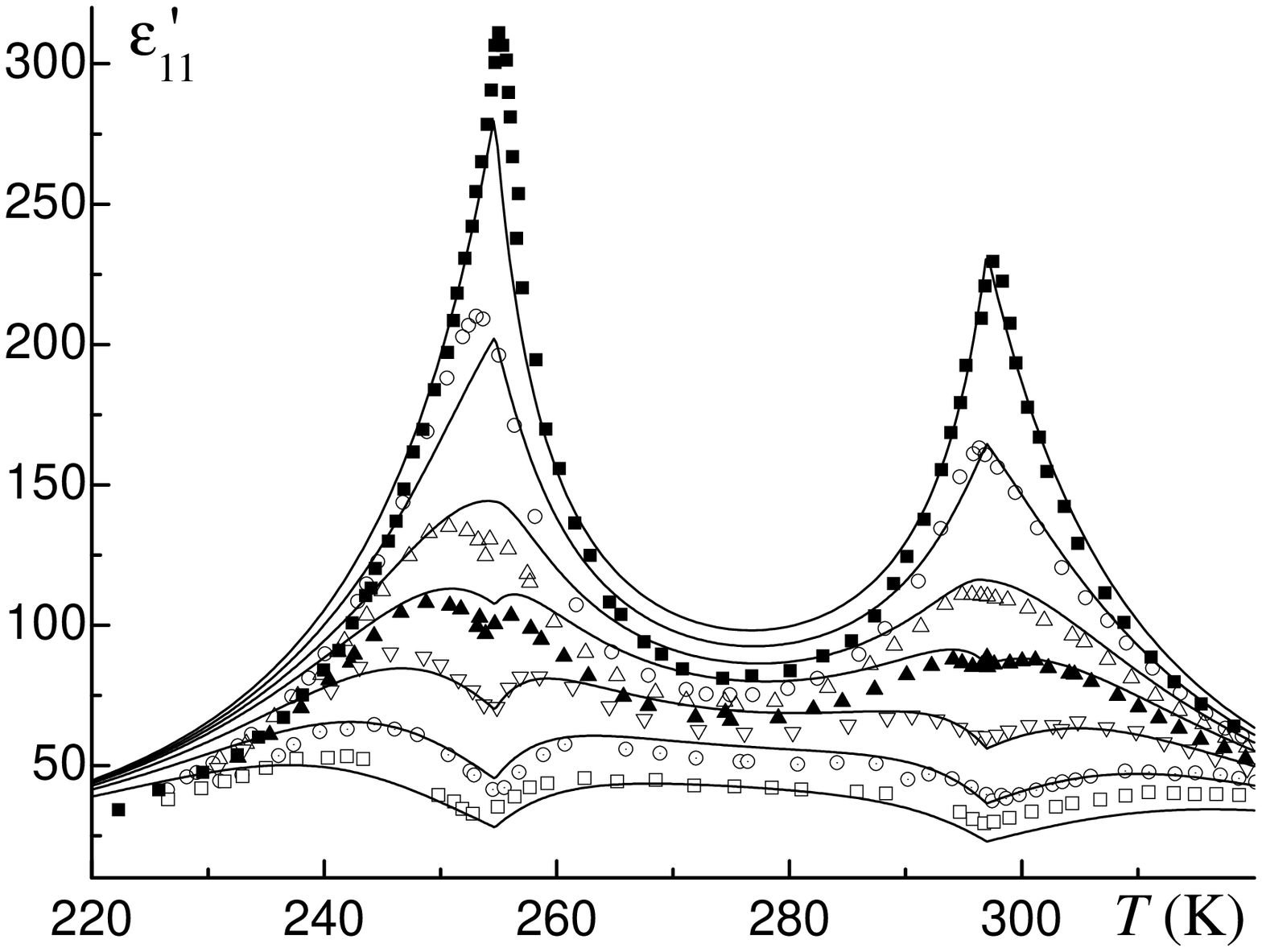}
\includegraphics[width=0.49\textwidth]{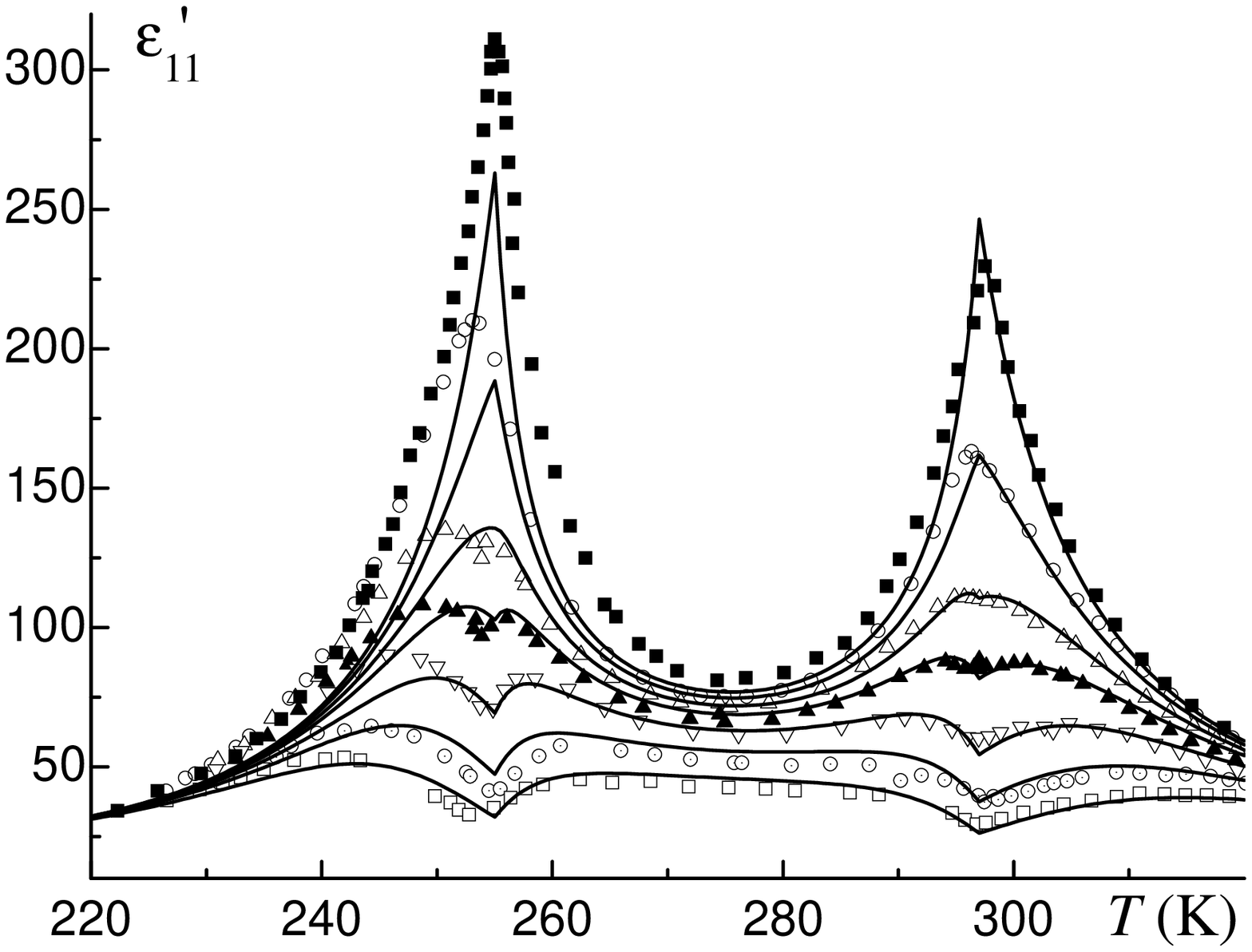}}
\caption{\label{epsre}The temperature dependence of the real part
of dynamic dielectric permittivity
 at different frequencies $\nu$
 (GHz): $\blacksquare$~--- 0.155,
  \protect\Large$\circ$\protect\small{}~--- 2.5,
    $\bigtriangleup$~--- 3.9,
 $\blacktriangle$~--- 5.1,
 $\bigtriangledown$~--- 7.05,
 $\odot$~--- 9.45,
 $\square$~--- 12.95. Solid lines are calculated within
the modifications of the Mitsui model without~\cite{ourrs} (left)
and  with~\cite{our-diagonal} (right) thermal strains; the symbols
are experimental points taken from~\cite{35}.}
 \end{figure}

\begin{figure}[hbt]
\centerline{\includegraphics[width=0.48\textwidth]{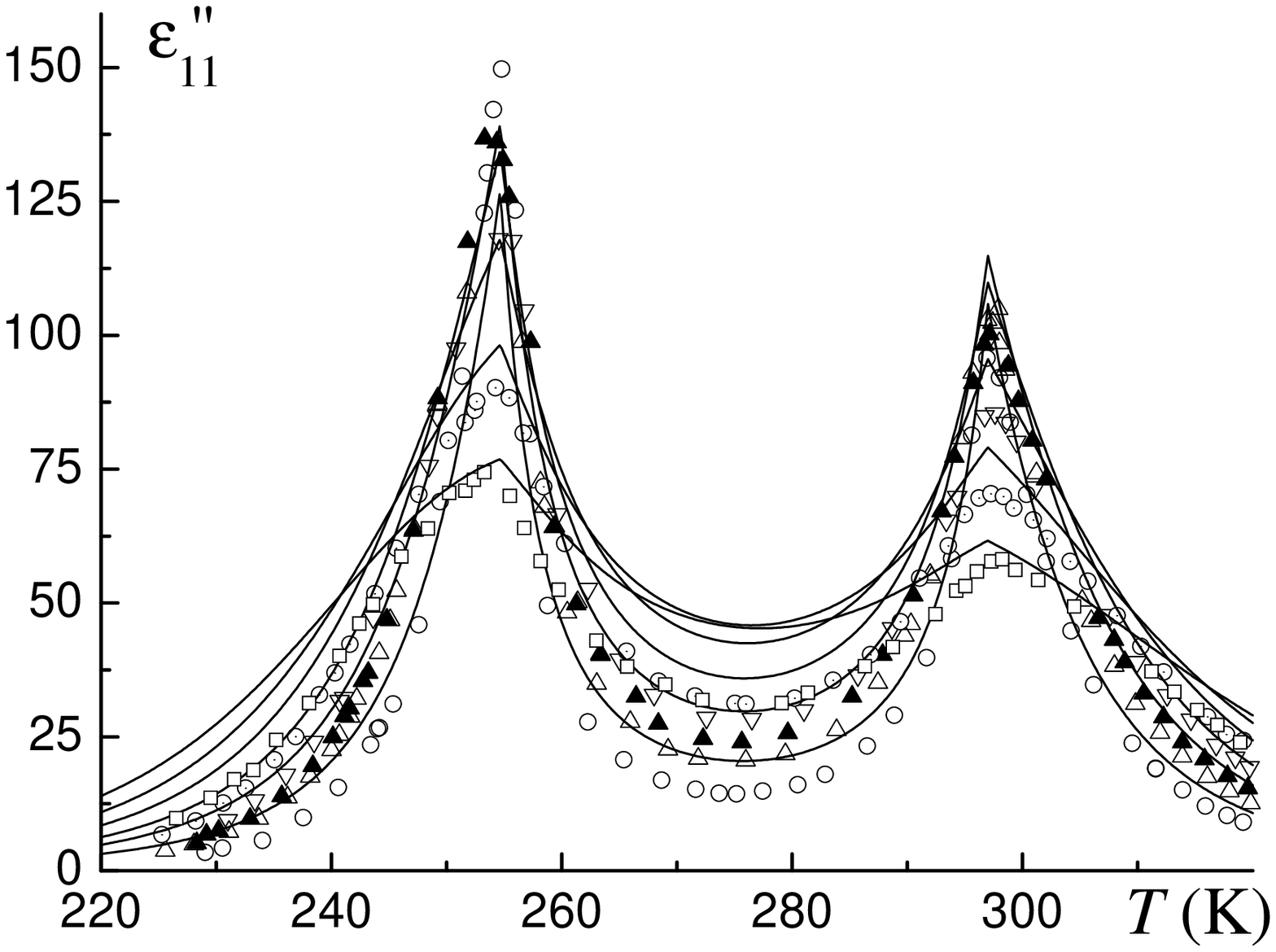}
\includegraphics[width=0.48\textwidth]{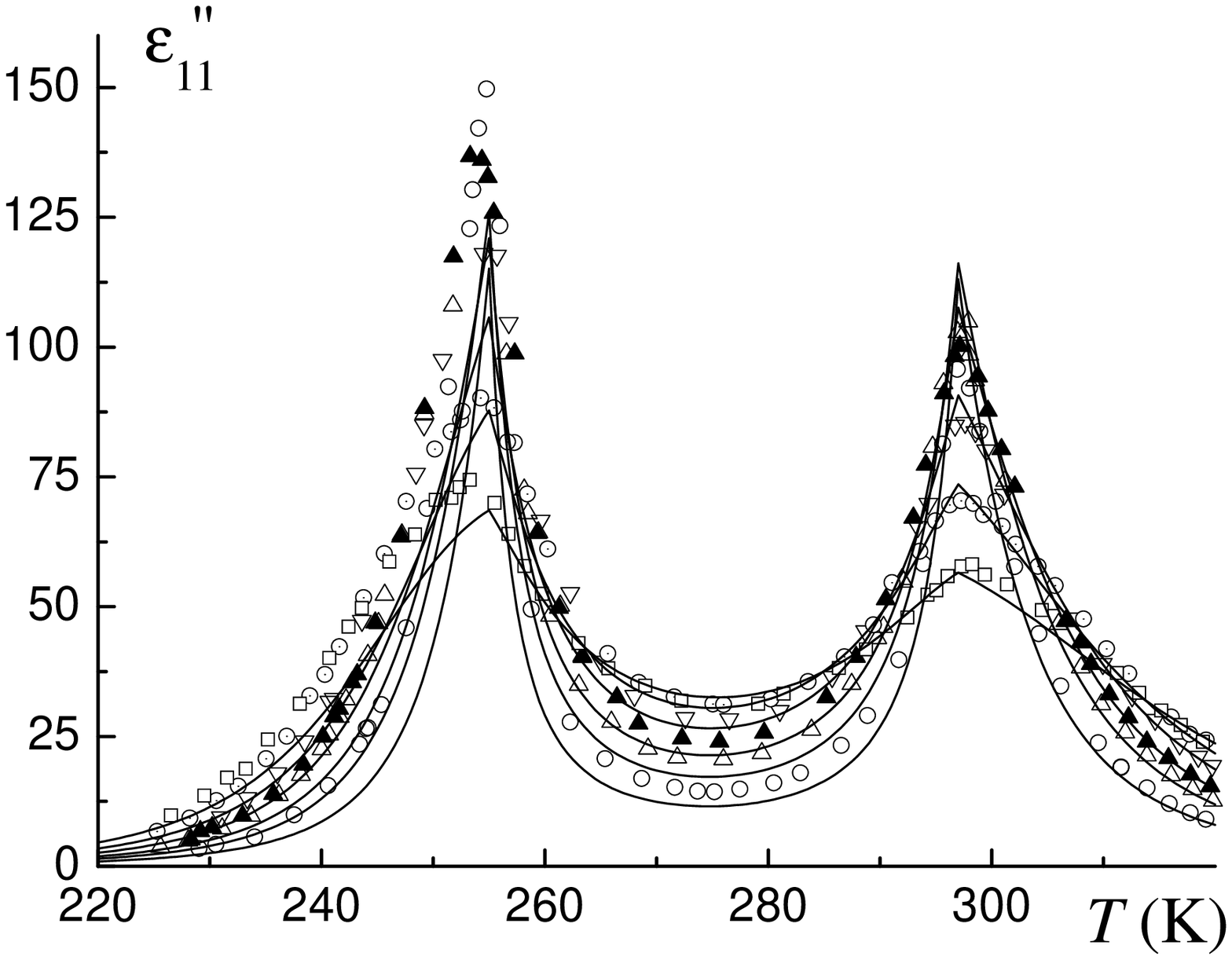}}
\caption{\label{epsim}The same for the imaginary part of the
permittivity.}
 \end{figure}

The temperature dependences of the real $\eps_{11}'$ and imaginary
$\eps_{11}''$ parts of the dynamic dielectric permittivity of
Rochelle salt calculated within the modifications of the Mitsui
model with~\cite{our-diagonal} and without~\cite{ourrs} thermal
strains are shown in figures~\ref{epsre} and \ref{epsim}.

Both models provide an excellent agreement with experimental data
\cite{35} for $\eps_{11}'$ in the upper paraelectric phase. For
$\eps_{11}''$, the newest results of the model~\cite{our-diagonal}
are slightly better, especially for very high frequencies (above
9~GHz).

The domain effects are switched off at microwave frequencies.
Therefore, they cannot explain the disagreement between the theory
and experiment in the ferroelectric phase for $\eps_{11}'$ and
$\eps_{11}''$, which took place in the previous modification of
the Mitsui model~\cite{ourrs}. Inclusion of thermal strains
allowed us to basically solve this problem; the obtained  fit using this
model is quite satisfactory in the ferroelectric phase. This
improvement has the same origin as that for spontaneous
polarization and spontaneous strain: the increased maximum value
of the order parameter $\xi$. This is the only exception when the
improvement for the permittivities could not be obtained by
changing the fitting procedure for $\mu_1$ and $\alpha$ without
taking into account the diagonal strains.

None of the models can properly describe the temperature
variation of  $\eps_{11}'$ or $\eps_{11}''$ below $T_{\rm C1}$. As
we have already mentioned, depending on the choice of  $\mu_1$ we
can obtain a good fit for $\eps_{11}'$ either between 240 and
255~K (as was done in~\cite{ourrs}) or below 230~K (in the model
with diagonal strains). The agreement with experiment for
$\eps_{11}''$  is only qualitatively correct at all temperatures
below $T_{\rm C1}$; quantitatively it is poor for both models.

\begin{figure}[hbt]
\centerline{\includegraphics[height=0.35\textwidth,width=0.4\textwidth]{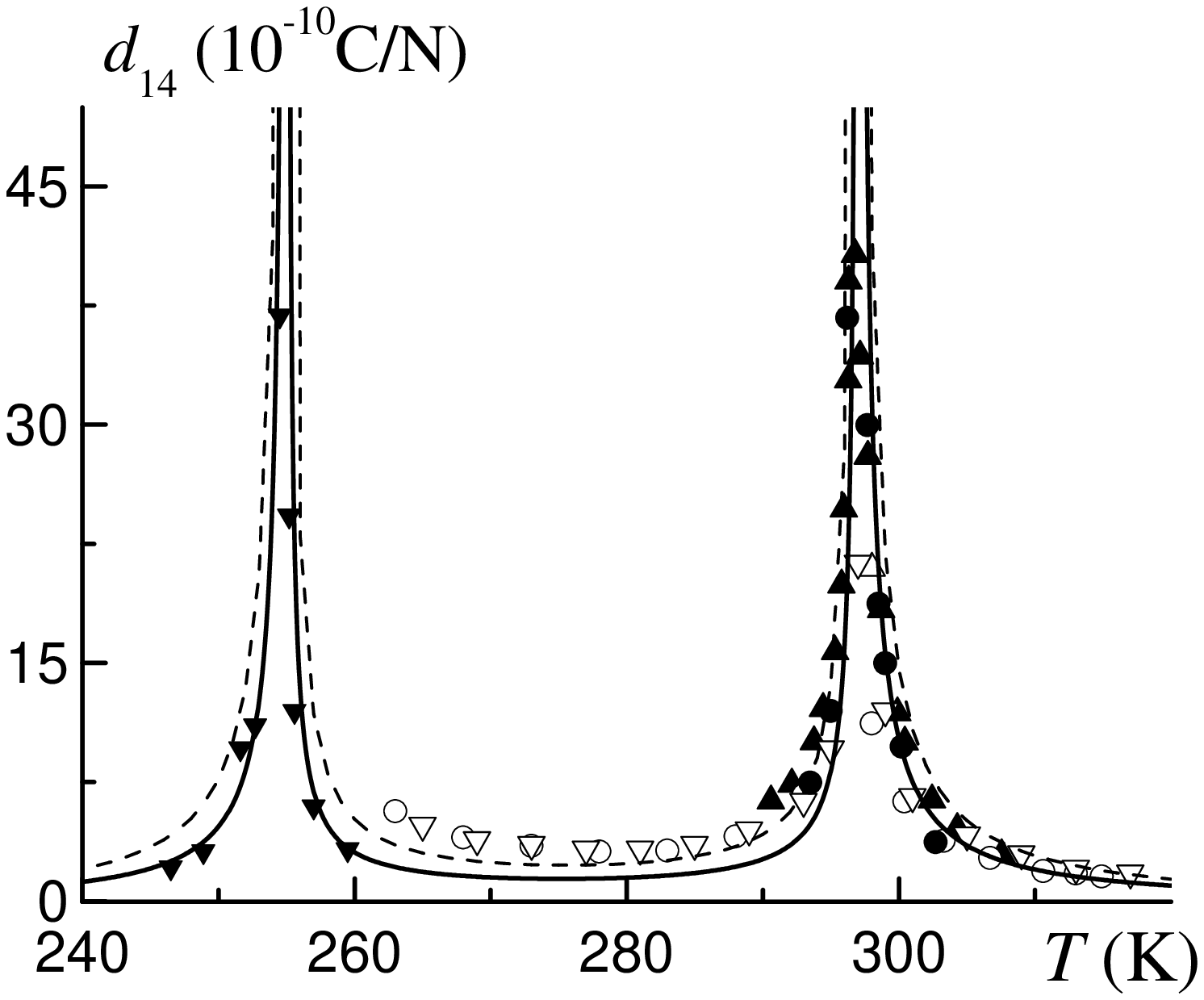}~~~~
\hspace{1cm}
\includegraphics[height=0.32\textwidth,width=0.4\textwidth]{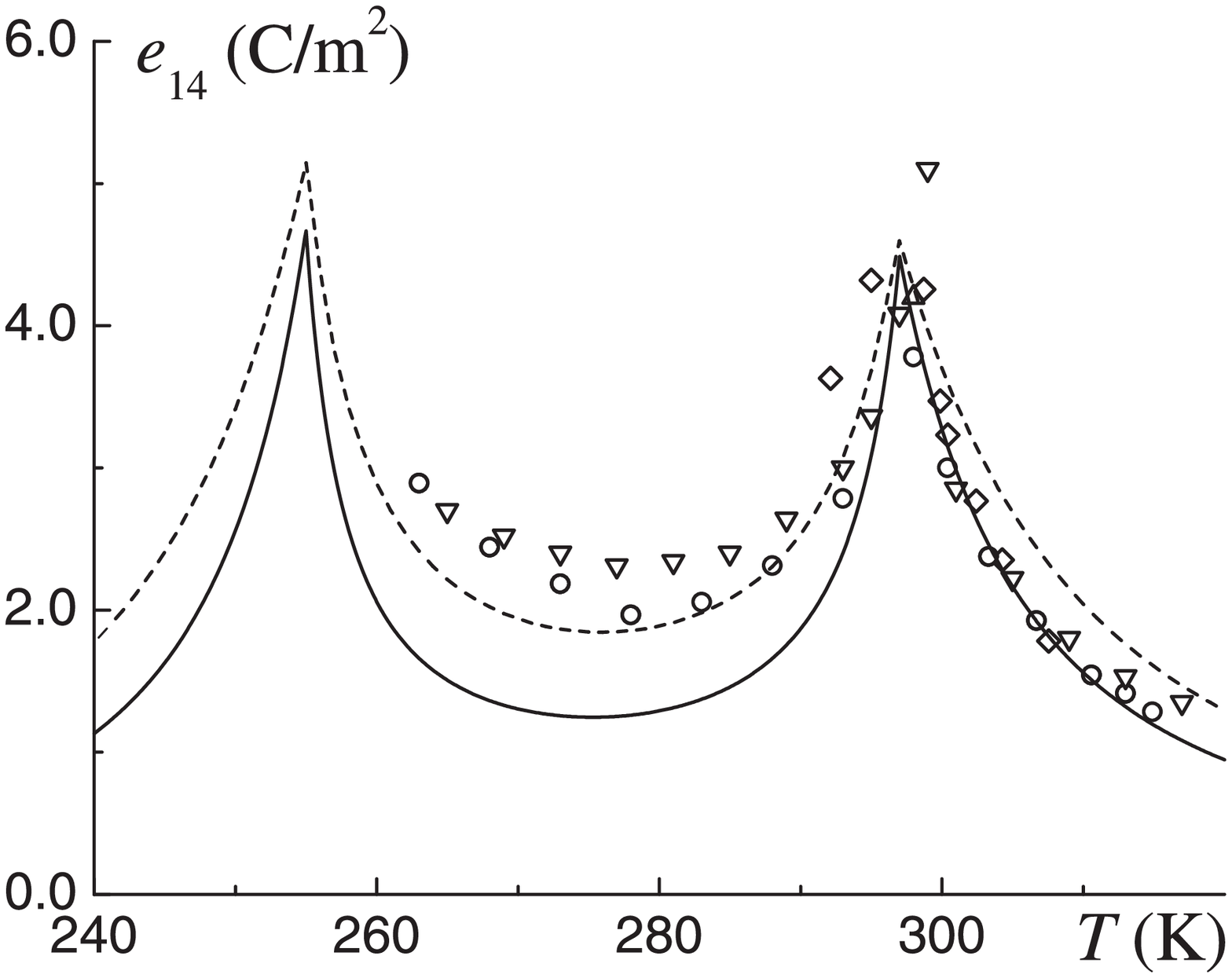} }
\caption{\label{e-d} Temperature dependences of piezoelectric
coefficients $d_{14}$ and $e_{14}$: $\bullet$~---~\cite{46},
$\blacktriangle$~---~\cite{45}, $\blacktriangledown$~---~\cite{47},
\Large$\circ$\small{}~---~\cite{24}, $\bigtriangledown$~---~\cite{33},
$\bigtriangleup$~---~\cite{43}, $\Diamond$~--- $e_{14}=d_{14}\cdot
c_{44}^E$~\cite{42,45}.
 Lines: the theory.
Solid line: the model of~\cite{our-diagonal}; dashed line:
\cite{ourrs}. Symbols: experimental points. }
\end{figure}

Comparison of the piezoelectric coefficients $d_{14}$ and $e_{14}$
calculated within the models~\cite{our-diagonal} and~\cite{ourrs}
is given in figure~\ref{e-d}. It can be seen that the agreement with
experiment in the lower paraelectric phase for $d_{14}$ and in the
upper paraelectric phase for $e_{14}$ obtained within the model
with the thermal strains is better.

\begin{figure}[hbt]
\centerline{\includegraphics[width=0.47\textwidth]{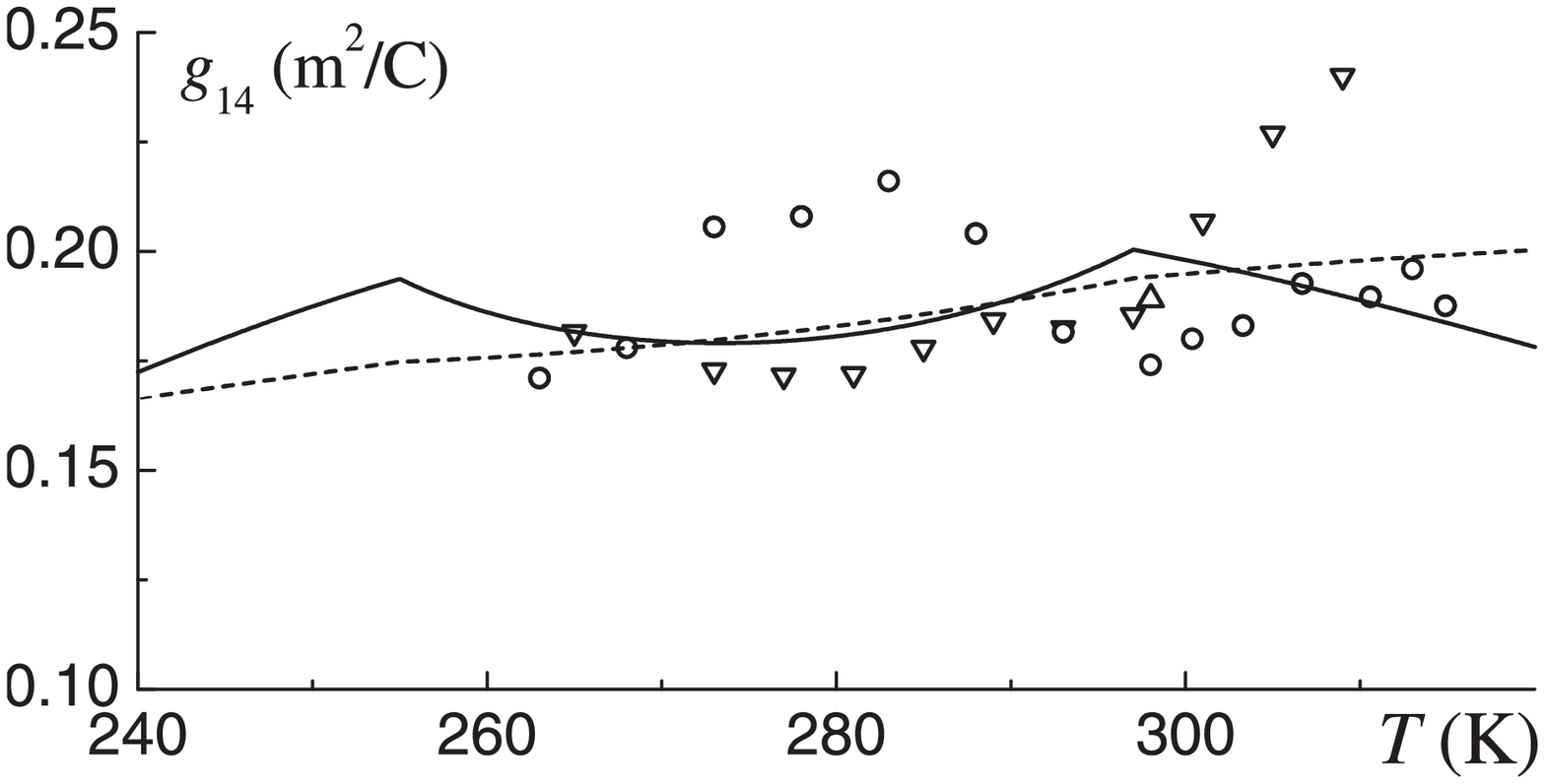}~~~
\includegraphics[width=0.47\textwidth]{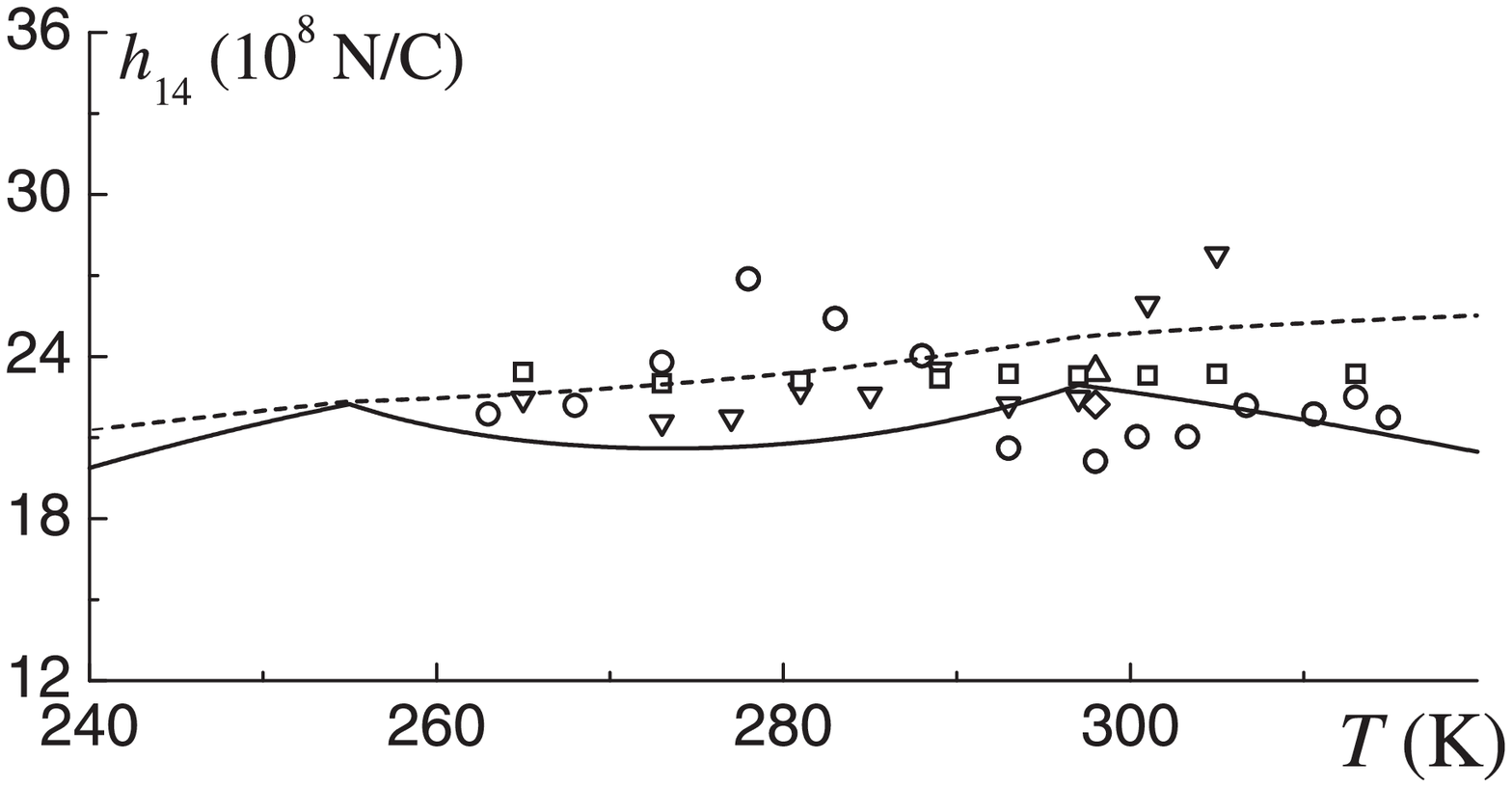} }
\caption{\label{g-h} Temperature dependences of piezoelectric
constants $g_{14}$ and $h_{14}$: $\square$~---~\cite{27},
$\bigtriangleup$~---~\cite{43}, \protect\Large$\circ$\protect\small{}~---~\cite{24}, $\bigtriangledown$~---
$h_{14}={e_{14}}/{\chi_{11}^{\varepsilon}}$ and
$g_{14}={d_{14}}/({\chi_{11}^{\varepsilon} + e_{14}d_{14}})$
\cite{33}, $\Diamond$~---~\cite{39}. Lines: the theory. Solid line:
the model of~\cite{our-diagonal}; dashed line:~\cite{ourrs}.
Symbols: experimental points. }
\end{figure}

Inclusion of diagonal strains enhances the temperature dependence
of the piezoelectric constants $g_{14}$ and $h_{14}$, as seen in
figure~\ref{g-h}. The theoretical curves, however, are still within
the dispersion range of the experimental data. This is the only
case when we could see that the diagonal strains
being taken into account produced some qualitative changes in the calculated
characteristics related to the strain $\eps_4$.
%
%
%
%
%

No direct effect of diagonal strains on the elastic constants
related to the shear strain $\eps_4$ is evident. Due to the
different value of the ``seed'' elastic constant $c_{44}^{E0}$ we
were able to improve the agreement with experiment for the elastic
constant at a constant field $c_{44}^{E}$ in the lower paraelectric
phase. The fit for the elastic constant at a constant polarization
$c_{44}^{P}$ becomes a little worse, though the theoretical curve
does not fall out of the dispersion range of the experimental
data.


\begin{figure}[hbt]
\centerline{\includegraphics[width=0.4\textwidth]{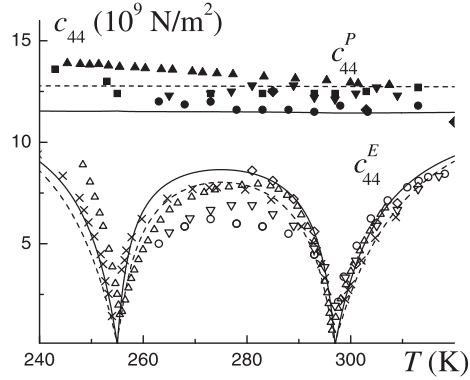}}
\caption{ Temperature dependence of elastic constant at constant
field $c_{44}^E$: $\times$~---~\cite{42}, $\bigtriangleup$~---
\cite{64B2}, $\Diamond$~---~\cite{40},
\protect\Large$\circ$\protect\small{}~--- ${1}/{s_{44}^E}$~\cite{24},
$\bigtriangledown$~--- ${1}/s_{44}^E$~\cite{33} and constant
polarization $c_{44}^P$: $\blacktriangle$~---~\cite{64B2},
$\blacksquare$~---~\cite{27}, $\blacklozenge$~---~\cite{39},
$\bullet$~--- $c_{44}^P={1}/{s_{44}^E} + e_{14}h_{14}$~\cite{24},
$\blacktriangledown$~--- ${1}/{s_{44}^E} +
e_{14}^2/\chi_{11}^{\varepsilon}$~\cite{33}. Lines: the theory.
Solid line: the model of~\cite{our-diagonal}; dashed line:
\cite{ourrs}. Symbols: experimental points.} \label{ce}
\end{figure}

It should be mentioned that the agreement with experiment for the
free static permittivity \linebreak (figure~\ref{fig-comp-chi}), piezoelectric
coefficient $d_{14}$ (figure~\ref{e-d}), and elastic constant
$c_{44}^E$ (figure~\ref{ce}) in the ferroelectric phase is getting
even worse when the diagonal strains are taken into account. The
disagreement, however, is obviously caused by the domain-wall
motion contributions into the mentioned characteristics, which are
not taken into account within the present~\cite{our-diagonal} or
previous~\cite{ourrs} versions of the Mitsui model. On the other
hand, the dynamic dielectric permittivity at microwave
frequencies (figures~\ref{epsre}, \ref{epsim}) and the free static
permittivity in high bias fields (above at least 0.5~kV/cm, see
next subsection), when the domain contributions are switched off,
are very well described by the Mitsui model with the diagonal
strains.

\subsection{Longitudinal electric field effects}
The longitudinal electric field $E_1$ directed along the axis of
spontaneous polarization, being the field conjugate to the order
parameter, smears out the phase transitions, decreases the maximal
values of permittivity $\varepsilon_{m}$, and shifts the upper
maximum temperature upward and the lower maximum temperature
downward. The phenomenological Landau-Devonshire formalism presumes
that $\varepsilon_{m}^{-1}$ and the shifts of permittivity maxima
temperatures $|\Delta T_{\mathrm{max}}(E)|$ are expected to vary with the
external field as $\sim E^{2/3}$~\cite{Lines}.

Those dependences are obtained starting with the simplified Landau
expansion of the thermodynamic potential (elastic and
piezoelectric contributions not considered)
\begin{equation}
\label{eq1} G(P_1) = G_0 + \frac{\alpha }{2}P_1^2 + \frac{\beta
}{4}P_1^4\,.
\end{equation}
From (\ref{eq1}), the equations for polarization and inverse
permittivity follow
\begin{equation}\label{polarization}
E_1=\left(\frac{\partial G}{\partial P_1}\right)=\alpha P_1 +
\beta P_1^3, \qquad \eps_{11}^{-1}=
\varepsilon_0\left(\frac{\partial E_1}{\partial
P_1}\right)=\varepsilon_0\left(\alpha  + 3 \beta P_1^2\right).
\end{equation}
In the case of Rochelle salt, the expansion (\ref{eq1}) can be
performed near each of the two transitions separately, assuming a
linear temperature dependence $\alpha = \alpha_{T1} (T_{\rm
C1}-T)$ for the lower transition and $\alpha = \alpha_{T2}
(T-T_{\rm C2})$ for the upper one. Then, the field dependences of
$\varepsilon_m$ and $\Delta T_{\mathrm{max}}$ can be presented as
\cite{Lines}:
\begin{equation}
\label{eq3} \varepsilon_m^{ - 1} =
\frac32(4\beta)^{1/3}\varepsilon_0 E_1^{2 / 3}=k_1E_1^{2 / 3}.
\end{equation}
\begin{equation}
\label{eq2} |\Delta T_{\mathrm{max},i}| =
\frac34\frac{(4\beta)^{1/3}}{\alpha_{Ti}}E_1^{2 / 3}= k_{2i}E_1^{2
/ 3}, \qquad i=1,2.
\end{equation}

The previous calculations~\cite{our-field} of the bias field
dependences of the static free permittivity of Rochelle salt
performed within the model without the diagonal strains
\cite{our-diagonal} have shown that the theory, though being
qualitatively correct, strongly overestimates the field effect on
temperature and magnitude of the upper permittivity maximum. In
fact, the calculated curves in the vicinity of $T_{\rm C2}$
coincided with the experimental points obtained in much lower
fields. On the other hand, the theory accorded well with
experiment in the vicinity of the lower Curie temperature.

To correctly describe the field dependences of the static
dielectric permittivity of Rochelle salt  at temperatures near
$T_{\rm C2}$, we had to use in our calculations the effective field
$E_{\rm eff}\sim0.7E_{\rm ext}$  at $T_{\rm C2}$, instead of the
values of the actually applied in the experiments field $E_{\rm
ext}$, whereas $E_{\rm eff}=E_{\rm ext}$ below $T_{\rm C1}$. An
assumption  has been made that the external field is partially
screened out due to the space-charge build-up at blocking
electrodes~\cite{our-field}.

\begin{figure}[th]
\centerline{\includegraphics[width=0.6\textwidth]{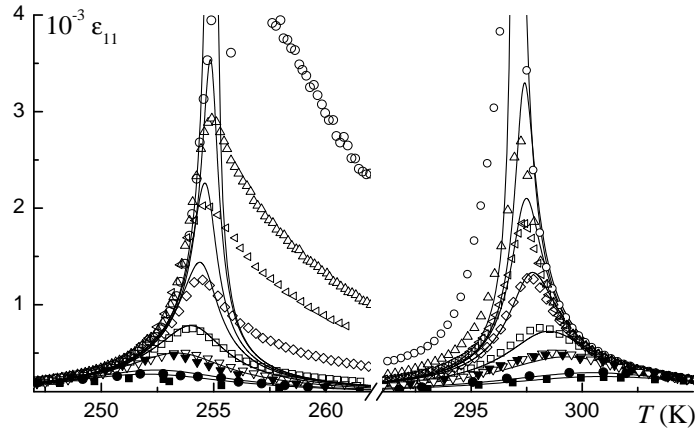}}
\caption{Temperature dependences of the dielectric permittivity of
Rochelle salt at different values of external electric field $E_1$
(kV/cm): {  \protect\Large$\circ$\protect\small{}}~--- 0, $\triangle$~--- 0.05, $\triangleleft$~---
0.1, $\lozenge$~--- 0.2, $\square$~--- 0.5, $\blacktriangledown$~--- 0.96,
$\triangledown$~--- 1.0, $\bullet$~--- 1.96,
$\blacksquare$~--- 2.46. Experimental points are taken from~\cite{ourexp} (open symbols) and~\cite{Akishige}  (closed
symbols). Lines: the theoretical results obtained within the model
of~\cite{our-diagonal}.} \label{eps-field}
\end{figure}
\begin{figure}[tbh]
\centerline{\includegraphics[width=0.95\textwidth]{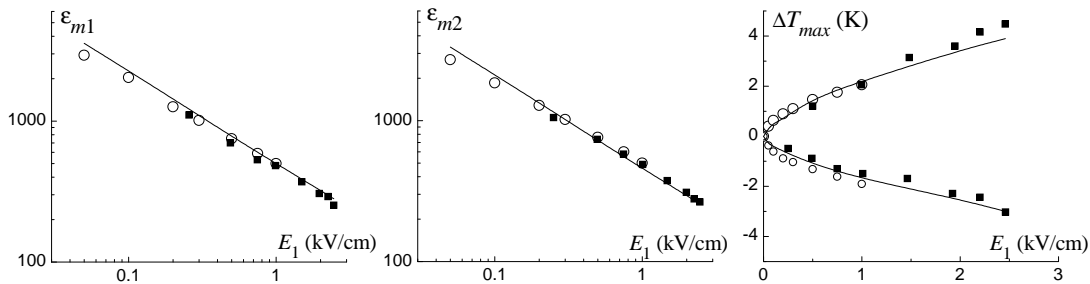}}
\caption{The field $E_1$ dependences of the magnitude and
temperature shifts of the dielectric permittivity maxima in
Rochelle salt. Experimental points are taken from~\cite{ourexp}
(open symbols) and~\cite{Akishige} (closed symbols).  Lines: the
theoretical results obtained within the model of
\cite{our-diagonal}.} \label{dtmax}
\end{figure}

The calculations performed within the modified Mitsui model with
the thermal strains~\cite{our-diagonal} show that this assumption
may be incorrect. Indeed, as seen in figures~\ref{eps-field} and
\ref{dtmax}, no overestimation of the field effects on the
temperature curves of the static free permittivity is obtained. In
fact, the experimental permittivity maxima are almost always lower
than the theoretical ones. The field dependences of the
temperature shifts of the maxima are rather well described by the
theory. The theoretical $\varepsilon_{m1,2}^{-1}$ and $|\Delta
T_{\mathrm{max}}|$  vary with the external field as $\sim E_1^{2/3}$, in
agreement with the Landau theory.

We can now deduce the values of the coefficients of the Landau
expansion. The theoretical curves in figure~\ref{dtmax} are well
approximated by (\ref{eq3}) and (\ref{eq2}) dependences with
\[
k_1 = 1.03 \cdot 10^{ -6}~{\rm (m/V)}^{2 / 3},\qquad
k_{21}=7.6\cdot 10^{ - 4}~{\rm K (m/V)}^{2 / 3},\qquad
k_{22}=9.9\cdot 10^{ - 4}~{\rm K (m/V)}^{2 / 3},
\]
yielding \[\beta =1.17\cdot 10^{14}~{\rm V~m}^{5}/{\rm
C}^{3},\qquad \alpha_{T1}=7.7\cdot 10^{7}~{\rm V m/(K C)},\qquad
\alpha_{T2}=5.8\cdot 10^{7}~{\rm V m/(K C)}.
\]

\begin{figure}[!t]
\centerline{\includegraphics[height=0.34\textwidth]{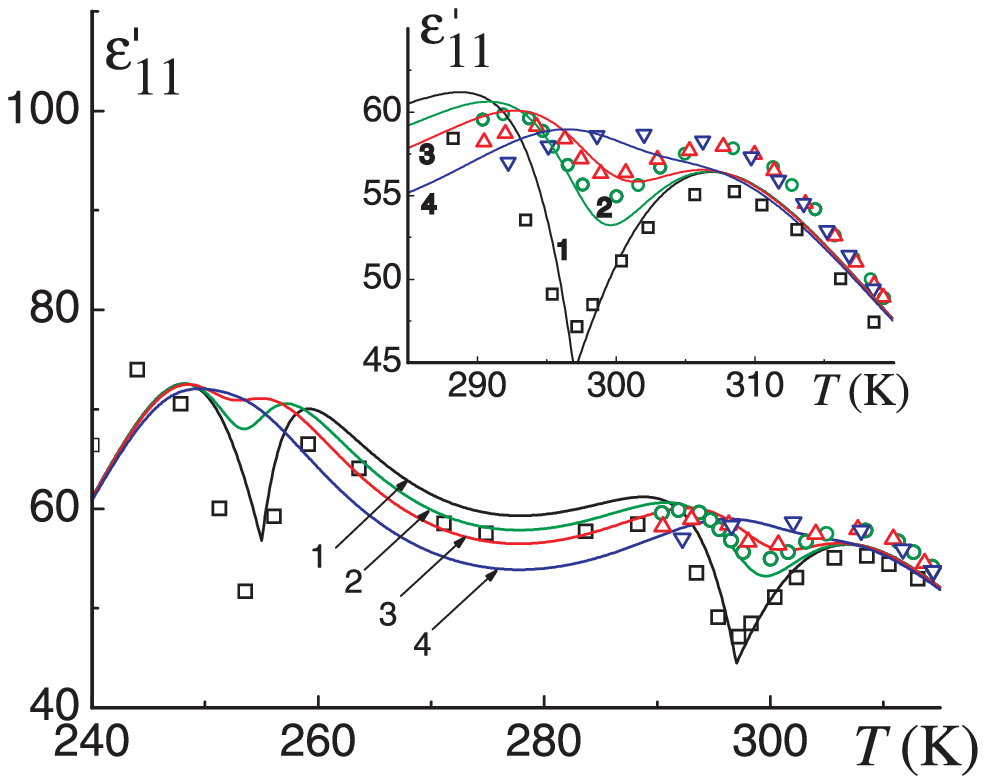}
\hfill 
\includegraphics[height=0.34\textwidth]{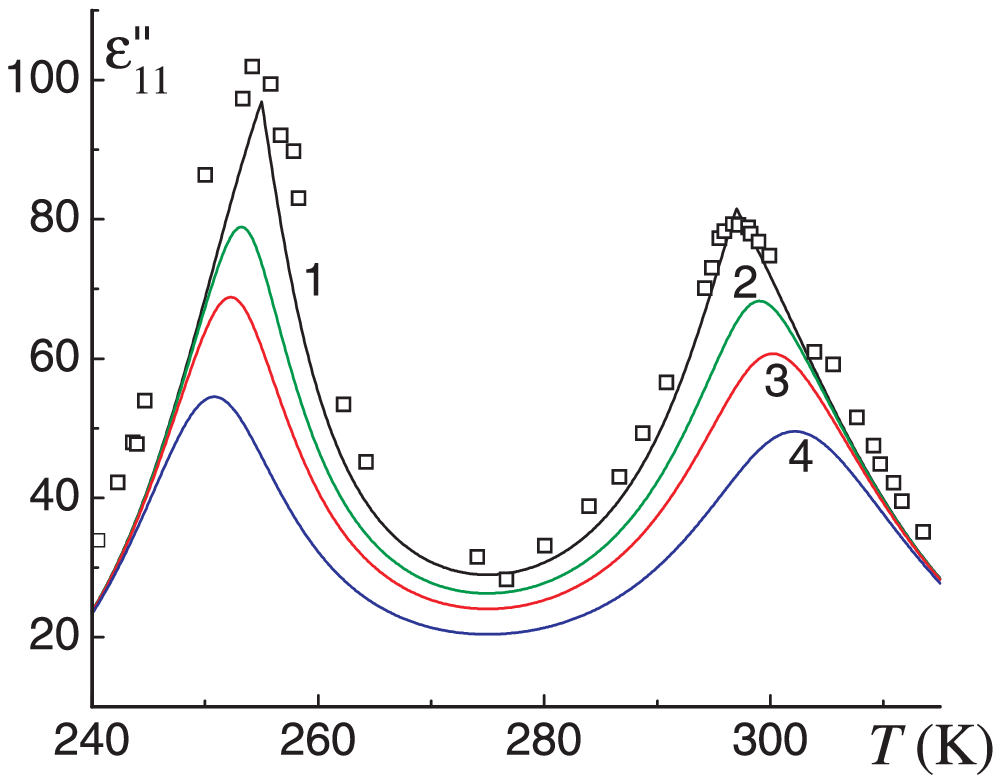}}
\caption{(Color online) Dynamic dielectric permittivity of Rochelle salt at
8.25~GHz at different values of the longitudinal field $E_{1}$
(kV/cm): 1, $\square$~--- 0; 2,   \protect\Large$\circ$\protect\small{}~--- 1, 3, $\triangle$~--- 2;
4, $\triangledown$~--- 4.1.   Lines: the theoretical results
obtained within the model of~\cite{our-diagonal}. Symbols:
experimental points taken from~\cite{35}. } \label{micro}
\end{figure}

Figure~\ref{micro} illustrates the field effects on the
microwave dynamic dielectric permittivity of Rochelle salt. The
theoretical curves were calculated within the model with the
thermal strains. In zero field, the real part of the permittivity
at 8.25~GHz has shallow minima at the transition points and a pair
of rounded maxima at both sides of each minimum~\cite{35}.
Application of d.c. bias smears out the minima, which shift toward
the corresponding paraelectric phases, and each pair of maxima
tends to coalesce. At sufficiently high fields, the maxima merge,
and the minima disappear. At temperatures far from the transition
regions, the real part of the permittivity only decreases with the
bias field. This behavior is illustrated in figure~\ref{micro}, and
it accords qualitatively with the experimental results of
\cite{35}. The behavior of the imaginary part of the permittivity is
qualitatively the same as that of the static permittivity: it
decreases with the field at all temperatures.

\section{Concluding remarks}
In the present paper we compare the results obtained within two
modifications (\cite{ourrs} and~\cite{our-diagonal}) of the
two-sublattice Mitsui model for physical characteristics of
Rochelle salt associated with the shear strain $\eps_4$. The role
of diagonal strains in the mentioned characteristics is explored.

It is shown that the thermal strains being taken into account bring about
mostly quantitative changes in the calculated
$\varepsilon_4$-strain-related characteristics. The only
qualitative effect is the enhancement of the temperature variation
of the piezoelectric constants $g_{14}$ and $h_{14}$.

The agreement with experiment for spontaneous polarization,
spontaneous strain $\eps_4$, and dynamic dielectric permittivity
in the ferroelectric phase, is significantly improved, as compared
to that of the modified Mitsui model without diagonal strains~\cite{ourrs}. The obtained improvement comes not from direct
contributions into these characteristics of diagonal strains, but from the fact that
the extended model yields larger values
of the order parameter $\xi$, resulting in a better fit to the
experiment.

For the elastic constants $c_{44}^{E,P}$ and dielectric
permittivities in the paraelectric phases, the inclusion of
diagonal strains did not result in any apparent quantitative
changes. We, however, adopted a slightly different route during
the fitting procedure for the parameters $\mu_1$, $c_{44}^{E0}$,
and $\alpha$, which allowed us to somewhat improve the agreement
with experiment. The same route can be taken within the model
without diagonal strains as well.

It has been also shown previously~\cite{our-diagonal} that the
model with diagonal strains  yields a better agreement with
experimental data for small anomalies of specific heat of
Rochelle salt at the Curie points than it was obtained with the
earlier model~\cite{ourrs}, especially for the magnitude of the
upper anomaly. In this case, in contrast to the above
discussed cases of spontaneous polarization and spontaneous strain, the
obtained improvement is caused by taking into account the explicit
contributions of diagonal strains.

We also reexamine the effect of the longitudinal  bias field
$E_1$ on the dielectric characteristics of Rochelle salt. The
model with thermal strains yields a fair agreement with
experiment; its results also accord with the prediction of the
Landau-Devonshire theory. In contrast to the calculations within
the model without thermal strains~\cite{our-field}, there is
no need to assume that the external electric field is
screened out by the space charge buildup at the blocking
electrodes.

\ukrainianpart

\title{Основні фізичні характеристики сеґнетової солі: вплив теплових деформацій}
\author{А.П. Моїна}
 \address{Інститут фізики конденсованих систем НАН України, вул. Свєнціцького, 1, 79011 Львів}

\makeukrtitle

\begin{abstract}
\tolerance=3000%
У статті порівнюються результати для пов'язаних зі зсувною деформацією
$\eps_4$ фізичних ха\-рак\-те\-рис\-тик сегнетової солі, отриманих в рамках нещодавно розвиненої
модифікованої двопідґраткової моделі Міцуї, що враховує деформацію $\eps_4$ та діагональні компоненти
тензора деформацій  $\eps_1$, $\eps_2$, $\eps_3$,  з результатами
попередньої модифікації моделі Міцуї, що враховує лише $\eps_4$.
В рамках моделі з діагональними (тепловими) деформаціями досліджено вплив поздовжнього електричного поля
$E_1$ на діелектричні властивості сеґнетової солі.

\keywords сеґнетова сіль, теплове розширення, модель Міцуї, деформації, електричне поле

\pacs 65.40.De, 77.65.Bn, 77.22.Gm, 77.22.Ch

\end{abstract}

\end{document}